\documentclass[epjST]{svjour}
\newcommand{\subparagraph}{}
\usepackage{graphicx}
\usepackage[utf8x]{inputenc}
\usepackage{amsmath}
\usepackage[colorinlistoftodos]{todonotes}
\usepackage{titlesec}
\usepackage{placeins}
\usepackage[labelfont=bf]{caption}
\usepackage{subcaption}
\usepackage{amssymb}
\usepackage{tikz}
\usepackage{xfrac}
\usepackage{nicefrac}
\usepackage{gensymb}
\usetikzlibrary{arrows}

\begin{document}
\title{Dynamic wetting failure in curtain coating by the Volume-of-Fluid method}
\subtitle{Volume-of-Fluid simulations on quadtree meshes}
\author{Tomas Fullana, Stéphane Zaleski \and Stéphane Popinet}
\institute{Sorbonne Université, CNRS, Institut Jean le Rond d'Alembert, UMR 7190, F-75005, Paris, France}
\abstract{In this paper we investigate dynamic wetting in the curtain coating configuration. The two-phase Navier-Stokes equations are solved by a Volume-of-Fluid method on an adaptive Cartesian mesh. We introduce the Navier boundary condition to regularize the solution at the triple point and remove the implicit numerical slip induced by the cell-centered interface advection. We use a constant contact angle to describe the dynamic contact line.
The resolution of the governing equations allows us to predict the substrate velocity at which wetting failure occurs. The model predictions are compared with prior computations of Liu et al. \cite{liu_vandre_carvalho_kumar_2016,Liu2018} and experimental observations of Blake et al. \cite{Blake1999} and Marston et al. \cite{Marston2009}. 
} 
\maketitle
\section{Introduction}
\label{intro}
\begin{figure}[h]
\begin{center}
\begin{tikzpicture}

\draw[very thick, -] (-1.5,4) -- (-1.5,1.5);
\draw[very thick, -] (-0.5,4) -- (-0.5,1.5);
\draw[very thick, -] (-1.5,1.5)  to[bend right=30] (-0.6,0);
\draw[very thick, -] (-0.5,1.5)  to[bend right=40] (0.8,0.2);
\draw[very thick, -] (0.8,0.2) -- (4,0.2);
\draw [fill=black!20,black!20] (-4,-0.5) rectangle (4,0); 
\draw [fill=blue!50,blue!50] (-1.5,4) -- (-1.5,1.5) to[bend right=30] (-0.6,0) -- (4,0) -- (4,0.2) -- (0.8,0.2) to[bend left=40] (-0.5,1.5) -- (-0.5,4) -- (-1.5,4);

\draw[thick, ->] (-1.25,4) -- (-1.25,3.5);
\draw[thick, ->] (-1.,4) -- (-1.,3.5);
\draw[thick, ->] (-0.75,4) -- (-0.75,3.5);

\draw[thick, dashed, <->] (2.5,4) -- (2.5,0);
\draw[thick, dashed, <->] (-1.5,4.2) -- (-0.5,4.2);

\draw[thick, ->] (0.,-0.25) -- (0.5,-0.25);

\draw[very thick,red!60, ->] (-1.05,0.1) to[bend left=40] (-2.75,1);


\draw[very thick, red!60] (-0.6,0) circle (0.4cm);


\draw [fill=black!20,black!20] (-3.5,1.5) rectangle (-2,1.2);
\draw [thick, ->] (-3,1.35) -- (-2.5,1.35);
\draw[thick, -] (-2.3, 1.5) to[bend right=20] (-2.6, 1.9);
\draw[thick, -] (-2.6, 1.9) --(-3.5, 2.5);
\draw [fill=blue!50,blue!50] (-2.3, 1.5) to[bend right=20] (-2.6, 1.9) --(-3.5, 2.5) -- (-2, 2.5) --(-2, 1.5);
\draw [ -] (-2.1, 1.5) to[bend right=40] (-2.4, 1.7);
\draw[thick, -] (-2.32, 1.5) --(-3.1, 2.2);
\draw[very thick] (-3.1,2.2) circle (0.03cm);

\draw (2.5,2) node[thick,right] {$h_{c}$};
\draw (-1,4.2) node[thick,above] {$d_{c}$};
\draw (0.5,-0.25) node[thick,right] {U};
\draw (-1, 3.5) node[thick,below] {V};
\draw (-1,2) node[thick] {$\rho_{l}, \mu_{l}$};
\draw (0.75,2) node[thick] {$\rho_{g}, \mu_{g}$};
\draw (-2.2,1.6) node[above] {$\theta_{m}$};
\draw (-3.1,2.2) node[left] {IP};

\end{tikzpicture}
\end{center}

\caption{Schematic of the curtain coating configuration. The system parameters are: $h_c$ the curtain height, $d_c$ the curtain width, $\rho_{l}$, $\rho_{g}$ and $\mu_{l}$, $\mu_{g}$ the densities and viscosities of the liquid phase and the gas phase respectively, $U$ the substrate velocity, $V$ the feed flow velocity and $\theta_m$ the imposed contact angle. The inflexion point noted IP corresponds to the point at which the curvature of the interface is zero. The distance from the triple point to the IP will be used as a control quantity in the validation section.}
\label{ghost2}
\end{figure}
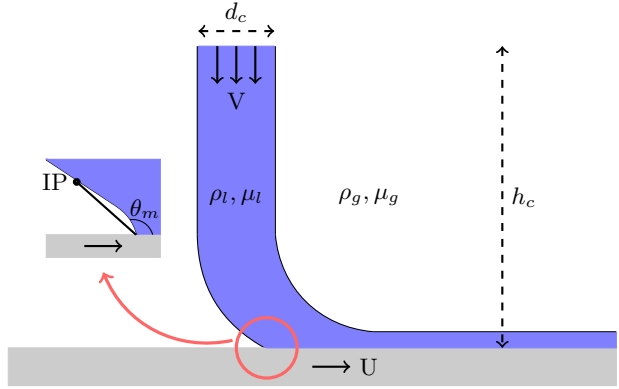

\begin{figure}[h]
  \centering
  \begin{subfigure}[h]{0.3\textwidth}
    \includegraphics[width=0.9\textwidth]{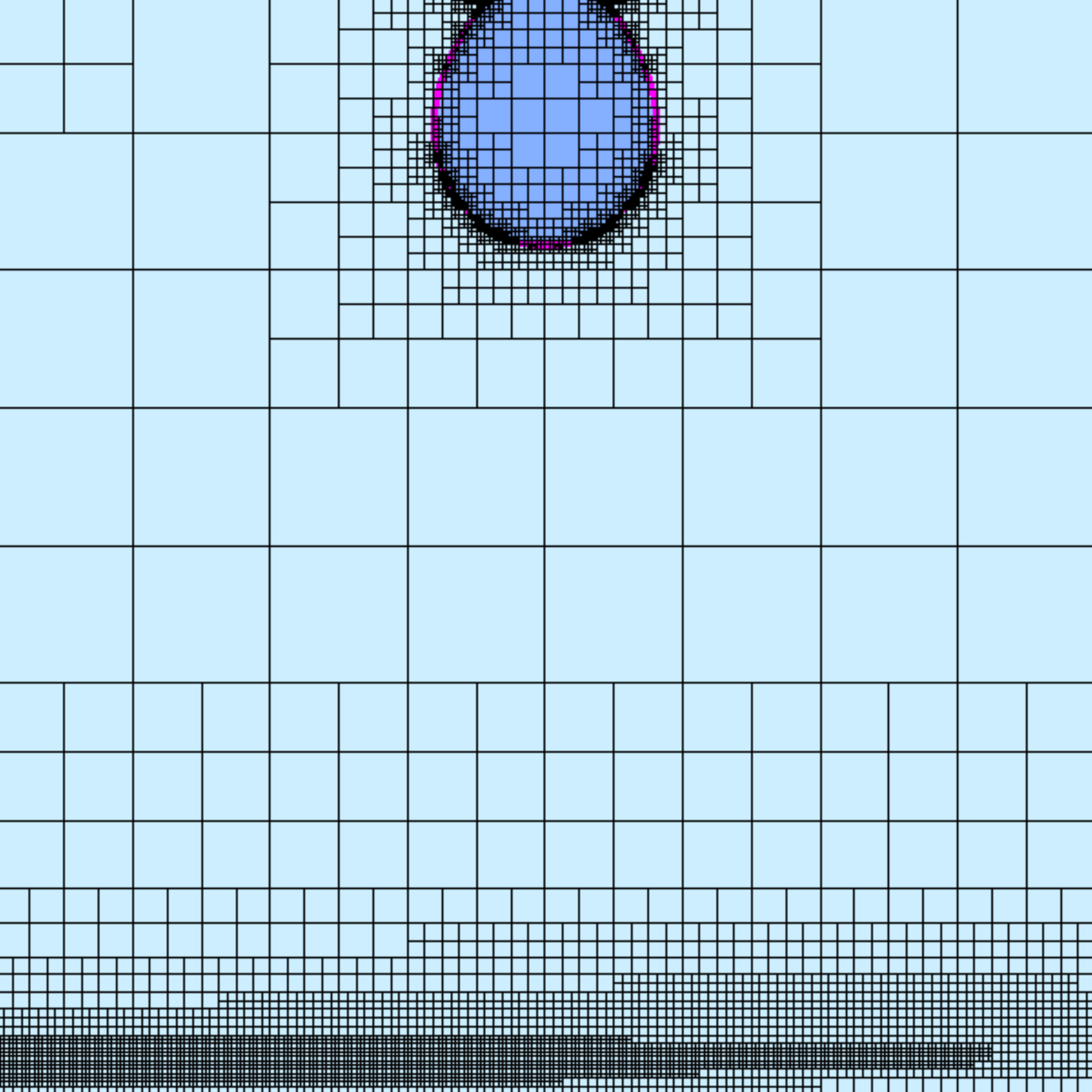}
    \caption*{$t^{\star} = 1.25$}
  \end{subfigure}
  \begin{subfigure}[h]{0.3\textwidth}
    \includegraphics[width=0.9\textwidth]{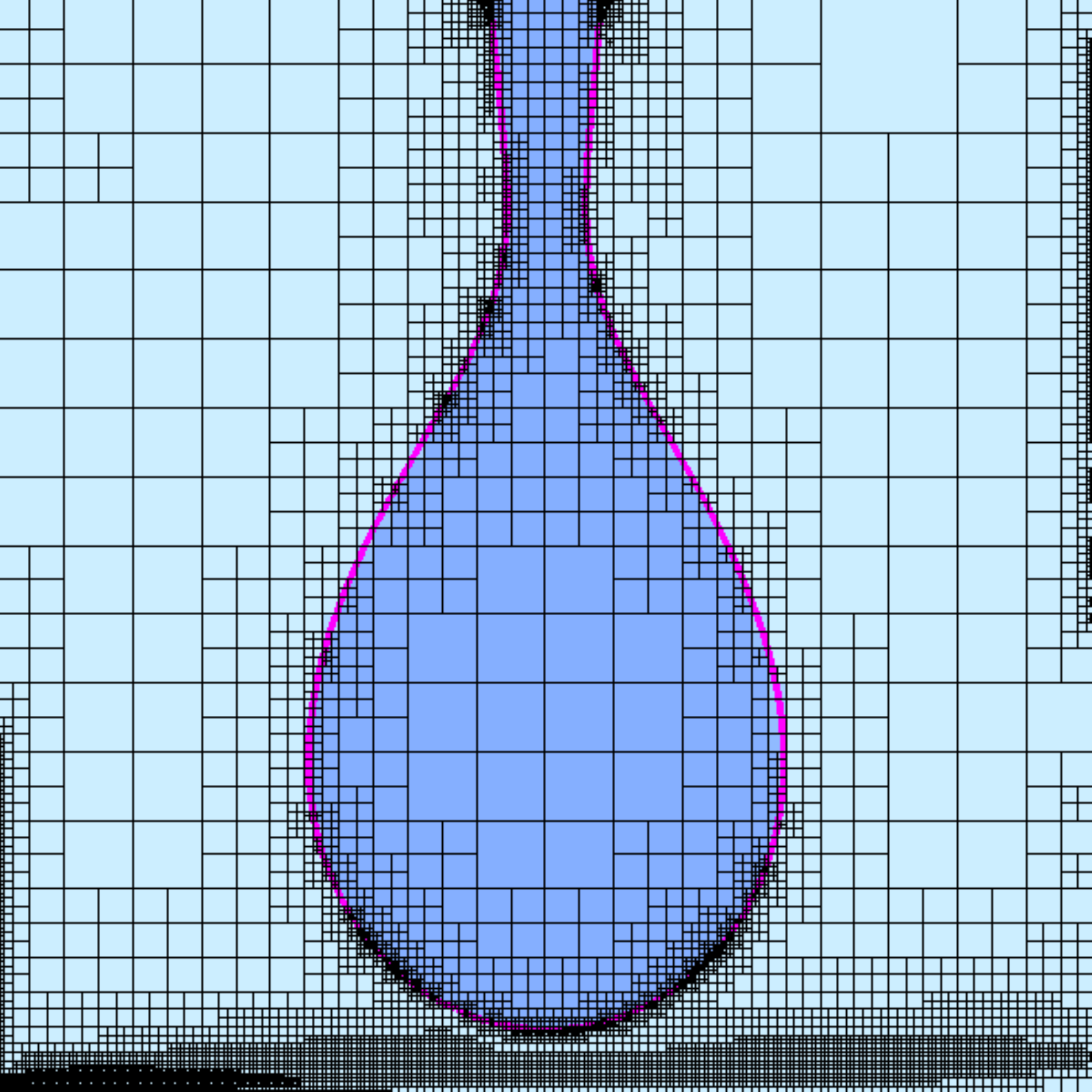}
    \caption*{$t^{\star} = 8.75$}
  \end{subfigure}
  \begin{subfigure}[h]{0.3\textwidth}
    \includegraphics[width=0.9\textwidth]{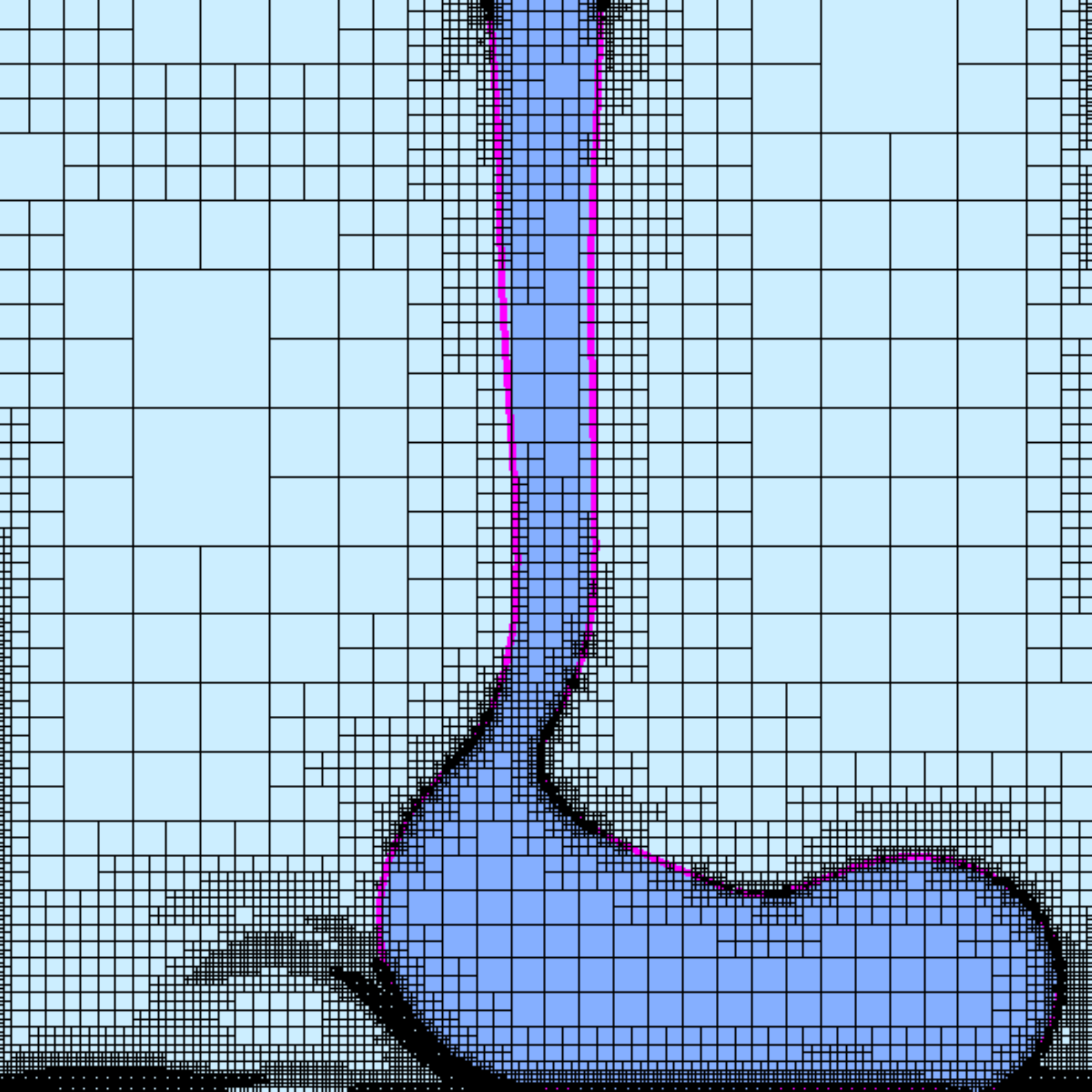}
    \caption*{$t^{\star} = 13.25$}
  \end{subfigure}
  \begin{subfigure}[h]{0.3\textwidth}
    \includegraphics[width=0.9\textwidth]{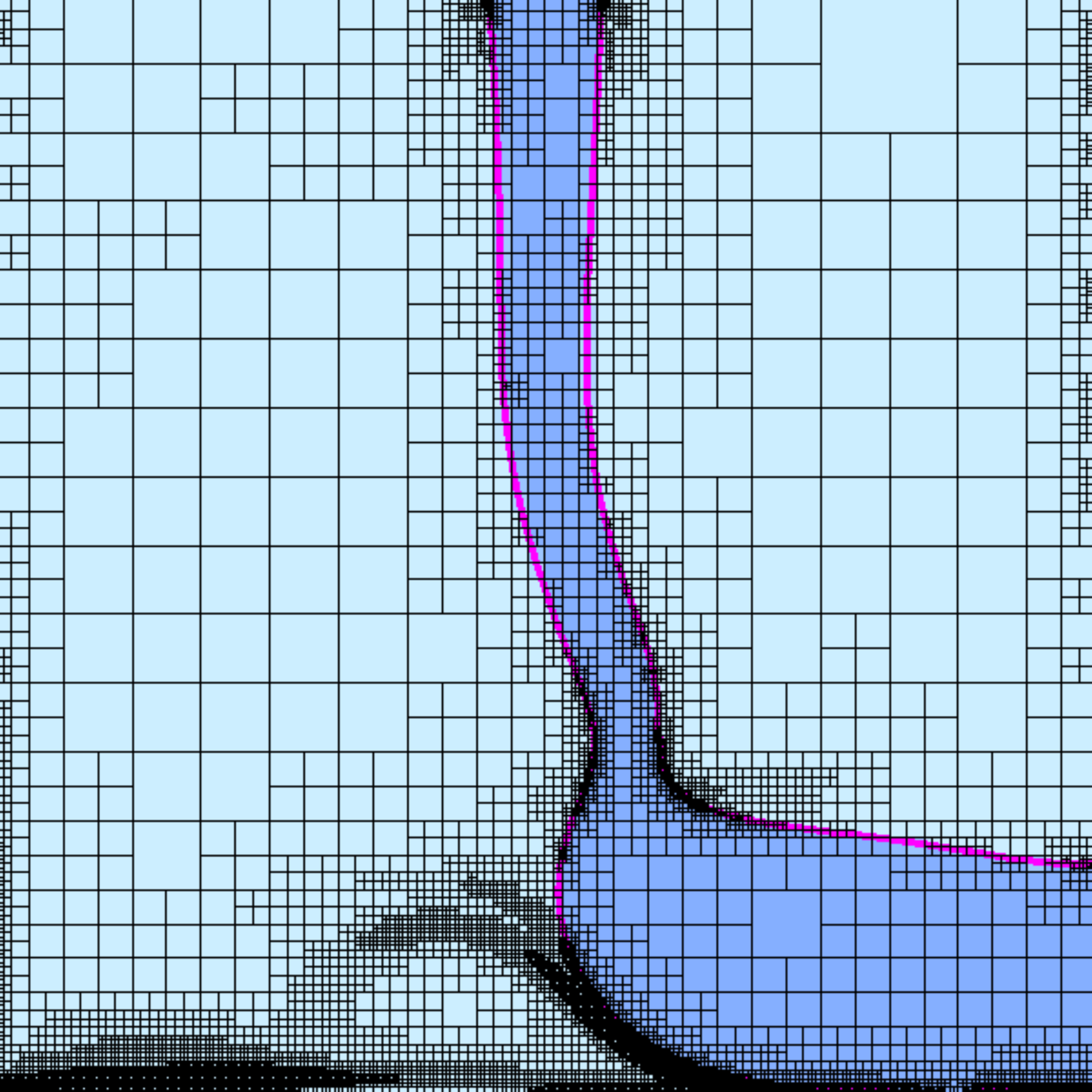}
    \caption*{$t^{\star} = 20$}
  \end{subfigure}
  \begin{subfigure}[h]{0.3\textwidth}
    \includegraphics[width=0.9\textwidth]{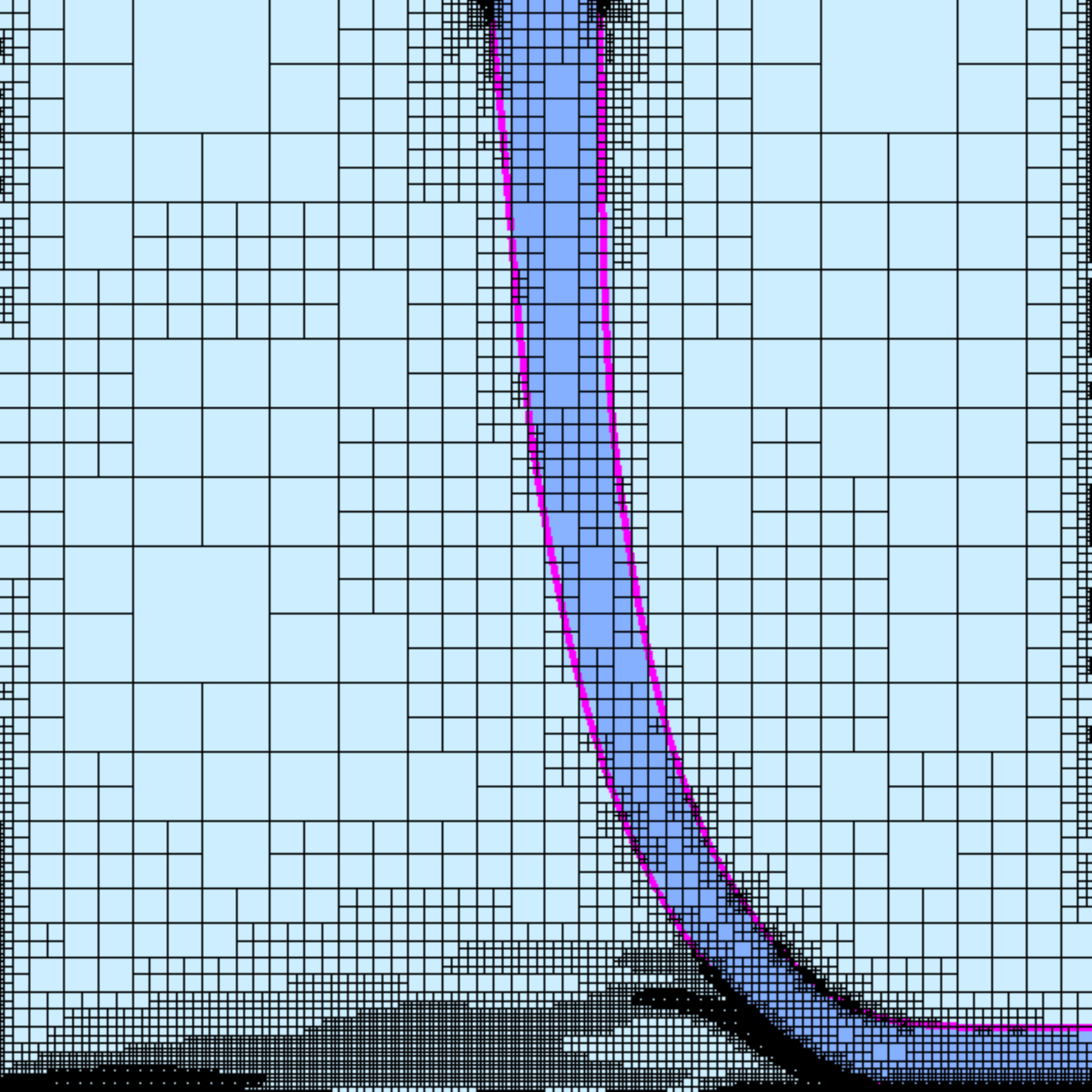}
    \caption*{$t^{\star} = 27.5$}
  \end{subfigure}
  \begin{subfigure}[h]{0.3\textwidth}
    \includegraphics[width=0.9\textwidth]{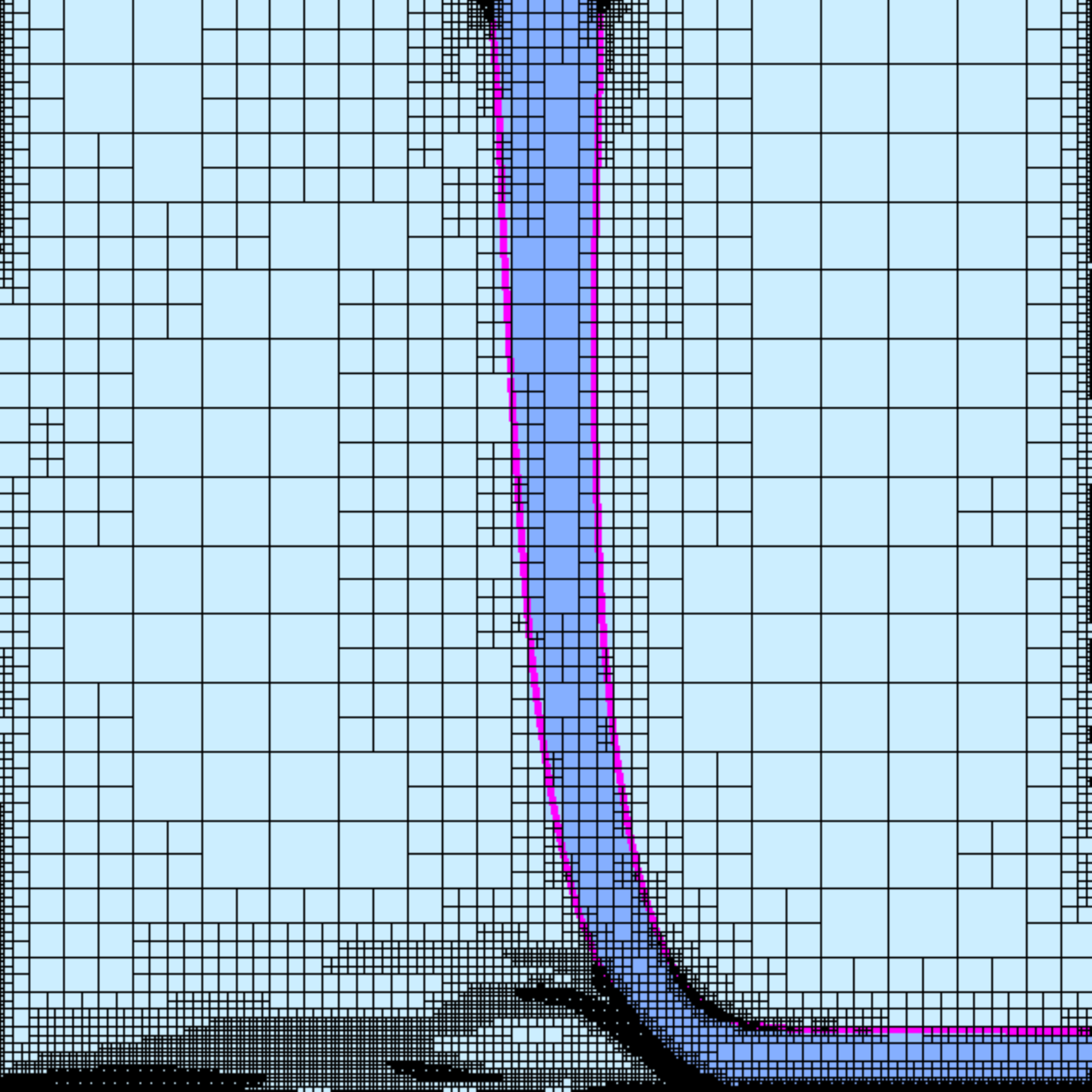}
    \caption*{$t^{\star} = 67.5$}
  \end{subfigure}
  \caption{Time series of a curtain coating simulation for Re = 12, Ca = 1.33. The dimensionless time $t^{\star}$ is scaled with the viscous time scale $T = \mu_l/\rho_l h_{c}^2$.}
  \label{time}
\end{figure}

The motion of the contact line poses, since Huh et al. \cite{HUH197185,huh_mason_1977}, a remarkable problem because of the contradiction between the no-slip condition on the substrate and the motion of the contact line. The slip length theory, expressed as a Navier boundary condition (NBC), is often used as a regularisation of the no-slip paradox at the triple point \cite{liu_vandre_carvalho_kumar_2016,Liu2018,Legendre2014}. In this paper, we will present a numerical model for the dynamic contact line that allows us to accurately represent the physics of the coating of a free surface. We will investigate different curtain coating configurations by carrying out two-dimensional Volume-of-Fluid (VOF) simulations. In the curtain coating system (\textbf{Figure \ref{ghost2}}), a liquid is falling with a velocity $V$ on a plate moving at velocity $U$. When the liquid reaches the solid substrate, it starts coating the free surface, as shown in the time series example (\textbf{Figure \ref{time}}). A steady-state solution is only obtained for given sets of physical parameters and the onset of wetting failure can be predicted by studying a range of capillary and Reynolds numbers by varying $U$ and $V$. We will make the same assumption as Liu et al. \cite{liu_vandre_carvalho_kumar_2016}: a constant contact angle coupled with a Navier boundary condition is sufficient to model the dynamic wetting system, provided that the air stresses are taken into account. In our model, these stresses are directly taken into account in the two-phase Navier-Stokes solver, the contact angle $\theta_m$ is imposed through height functions \cite{Afkhami2009,Afkhami2008} and a Navier boundary condition is implemented. The multi-scale nature of the curtain coating configuration can lead to numerical difficulties, in particular on the resolution of the smallest length scale, the slip length $\lambda$. The regularization of the contact line paradox is directly dependent on the accuracy of the solution near the contact line. We will study the convergence of the solution of the Liu et al. \cite{liu_vandre_carvalho_kumar_2016} configuration as we increase the number of grid points per slip length $\lambda / \Delta$. The system will then be extended to compare with experiments of Blake et al. \cite{Blake1999} and Marston et al. \cite{Marston2009}.


\section{Model}

The VOF method for representing fluid interfaces coupled with a flow solver is well-known to be suited for solving interfacial flows \cite{Scardovelli1999,Popinet1999}. In our study of the curtain coating system, we use the free software \emph{Basilisk}, a platform for the solution of partial differential equations on adaptive Cartesian meshes, developed by one of us \cite{Afkhami2009,Afkhami2008,Popinet2009,Popinet2015,Popinet2018,Afkhami2017}. \\
We consider the incompressible Navier-Stokes equations with variable density and surface tension:
$$\begin{array}{c}{\rho\left(\partial_{t} \mathbf{u}+\mathbf{u} \cdot \mathbf{\nabla} \mathbf{u}\right)=-\mathbf{\nabla} p+\mathbf{\nabla} \cdot(2 \mu \mathbf{D})+\sigma \kappa \delta_{s} \mathbf{n}} + \mathbf{g} \\ {\partial_{t} \rho+\mathbf{\nabla} \cdot(\rho \mathbf{u})=0} \\ {\mathbf{\nabla} \cdot \mathbf{u}=0}\end{array}$$
with $\mathbf{u}$ the fluid velocity, $\rho$ the fluid density, $\mu$ the fluid viscosity and $\mathbf{D}$ the deformation tensor defined as $D_{ij} = (\partial_{i}u_{j} + \partial_{j}u_{i})/2$, $\sigma$ the surface tension coefficient, $\kappa$ the curvature, $\delta_{s}$ the Dirac distribution function used for the sharp interface model, $\mathbf{n}$ the normal to the interface and $\mathbf{g}$ the acceleration of gravity.

For a two-phase flow, the volume fraction $c(\mathbf{x}, t)$ is defined as the integral of the first fluid's characteristic function in the control volume. The volume fraction $c(\mathbf{x}, t)$ is used to define the density and viscosity in the control volume:
 $$\begin{aligned} \rho({c}) & \equiv {c} \rho_{l}+(1-{c}) \rho_{g} \\ \mu({c}) & \equiv {c} \mu_{l}+(1-{c}) \mu_{g} \end{aligned}$$
with $\rho_{l}$, $\rho_{g}$ and $\mu_{l}$, $\mu_{g}$ the densities and viscosities of the liquid phase and the gas phase respectively.\\
The advection equation for the density is then replaced by the equation for the volume fraction:
$$\partial_{t} c+\mathbf{\nabla} \cdot(c \mathbf{u})=0$$
\\
The projection method is used to solve the incompressible Navier-Stokes equations combined with a Bell-Collela-Glaz advection scheme and a VOF method for interface tracking. For more details on the Navier-Stokes solver, see appendix-\ref{app}.\\

\begin{figure}[h]
\begin{center}
\begin{tikzpicture}

\draw [thin] (0,0) -- (7,0);
\draw [very thick] (0,1) -- (7,1);
\draw [very thick] (0,1) -- (7,1);

\draw [thin] (0,2) -- (7,2);
\draw [thin] (0,3) -- (7,3);

\draw [thin] (1,0) -- (1,4);
\draw [thin] (2,0) -- (2,4);
\draw [thin] (3,0) -- (3,4);
\draw [thin] (4,0) -- (4,4);
\draw [thin] (5,0) -- (5,4);
\draw [thin] (6,0) -- (6,4);

\draw [<->, thick] (1.9,0.5) -- (6,0.5);
\draw [<->, thick] (2.25,1.5) -- (6,1.5);
\draw [<->, thick] (2.85,2.5) -- (6,2.5);

\draw (6,0.5) node[thick,right] {$h_{0}$};
\draw (6,1.5) node[thick,right] {$h_{1}$};
\draw (6,2.5) node[thick,right] {$h_{2}$};

\draw [thick,dotted] (1.9,0) -- (1.9,1);
\draw [thick,dotted] (2.25,1) -- (2.25,2);
\draw [thick,dotted] (2.85,2) -- (2.85,3);

\draw [very thick] (2.05,1) -- (2.45,2);
\draw [very thick] (2.475,2) -- (3,2.675);
\draw [very thick] (3,2.7) -- (3.5,3);
\draw [very thick] (3.525,3) -- (4,3.2);
\draw [very thick] (4,3.225) -- (5,3.6);
\draw [very thick] (5,3.625) -- (6,3.8);

\draw [dashed, ->] (-0.5,1) -- (7.5,1);
\draw [dashed, ->] (0,0.5) -- (0,3.5);
\draw (-0.05,3.4) node[left] {$y$};
\draw (7.4,0.95) node[below] {$x$};

\end{tikzpicture}
\end{center}

\caption[Schematic of velocity profiles for no-slip and slip conditions and the ghost boundary layer. ]{Construction of the 2D height-functions near the contact line \cite{Afkhami2008}.}
\label{width}
\end{figure}
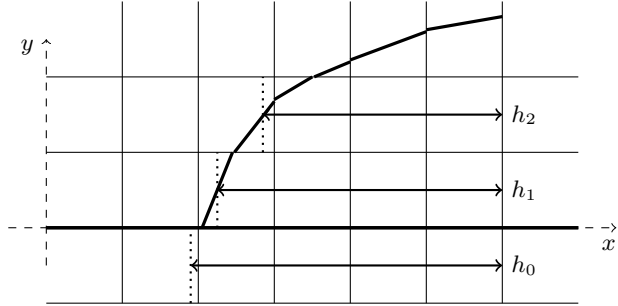

The resolution of the surface tension term is directly dependent on the accuracy of the curvature calculation. The Height-Function methodology is a VOF-based technique for calculating interface normals and curvatures \cite{Afkhami2009,Afkhami2008}. About each interface cell, fluid ‘heights’ are calculated by summing fluid volume in the grid direction closest to the normal of the interface. In two dimensions, a 7×3 stencil around an interface cell is constructed and the heights are evaluated by summing volume fractions horizontally (\textbf{Figure \ref{width}}): $$h_{j}=\sum_{k=i-3}^{k=i+3} c_{j, k} \Delta$$
with c the volume fraction and $\Delta$ the grid spacing. The heights are then used to compute the the interface normal $\mathbf{n}$ and the curvature $\kappa$: 
$$\begin{array}{c}{\mathbf{n}=\left(h_{x},-1\right)} \\  \\ {\kappa=\dfrac{h_{x x}}{\left(1+h_{x}^{2}\right)^{3 / 2}}}\end{array}$$
where $h_{x}$ and $h_{xx}$ are discretised using second-order central differences.


The dynamic contact line introduces a paradox at the triple point, where the no-slip boundary condition or Dirichlet boundary condition at the solid interface induces a non-integrability of the solution. Nevertheless, in the VOF method such a contradiction does not take place as the volume fraction is advected using the velocity half-a-cell away from the wall. In their paper, the authors of \cite{Afkhami2017}, showed that there exists a ‘numerical slip’ that is mesh dependent. In order to control such slip, it is useful to introduce a Navier boundary condition to explicitly define a physical slip length $\lambda$, that will be used as a fitting parameter in our simulations \cite{Legendre2014,Sui2014}. 
The NBC corresponding to a slip model in a 2D configuration with the substrate at $y = 0$ can be written as follows:

\begin{center}
\[\arraycolsep=4pt\def\arraystretch{2}
\begin{array}{rcl} 
\mathbf{u} -\lambda\dfrac{\partial\mathbf{u}}{\partial y} & =& \mathbf{U} \\
\mathbf{v} \cdot \mathbf{n} &=& 0
\end{array}\]
\end{center}{}


\noindent
with $\mathbf{u}$ and $\mathbf{v}$ the $x$ and $y$ component of the velocity at the solid boundary, $\mathbf{n}$ the normal to the wall, $\lambda$ the slip length, and $\mathbf{U}$  the prescribed velocity of the moving substrate. For more details on the numerical implementation of the NBC, we refer the reader to L\={a}cis et al. \cite{lacis}.\\
It is important to note that a numerical specification of the contact angle affects the overall flow calculation in two ways: it defines the orientation of the VOF reconstruction in cells that contain the contact line and it influences the calculation of the surface tension term by affecting the curvature computed in cells at and near the contact line. The orientation of the interface, characterized by the contact angle -- the angle between the normal to the interface at the contact line and the normal to the solid boundary -- is imposed in the contact line cell.

In Section \ref{sec3}, we will show that the Navier boundary condition coupled with a constant contact angle is sufficient to model the contact line motion in a curtain coating system and to reproduce the non-monotonic behaviour of the critical velocity as the liquid flow-rate increases.
\section{Validation}
\begin{figure}[h]
  \centering
  \begin{subfigure}[h]{0.492\textwidth}
    \includegraphics[width=0.9\textwidth]{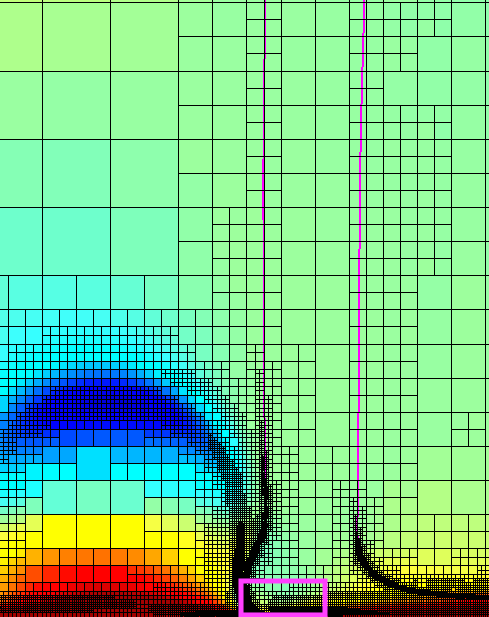}
  \end{subfigure}
  \begin{subfigure}[h]{0.45\textwidth}
  \includegraphics[width=0.75\textwidth]{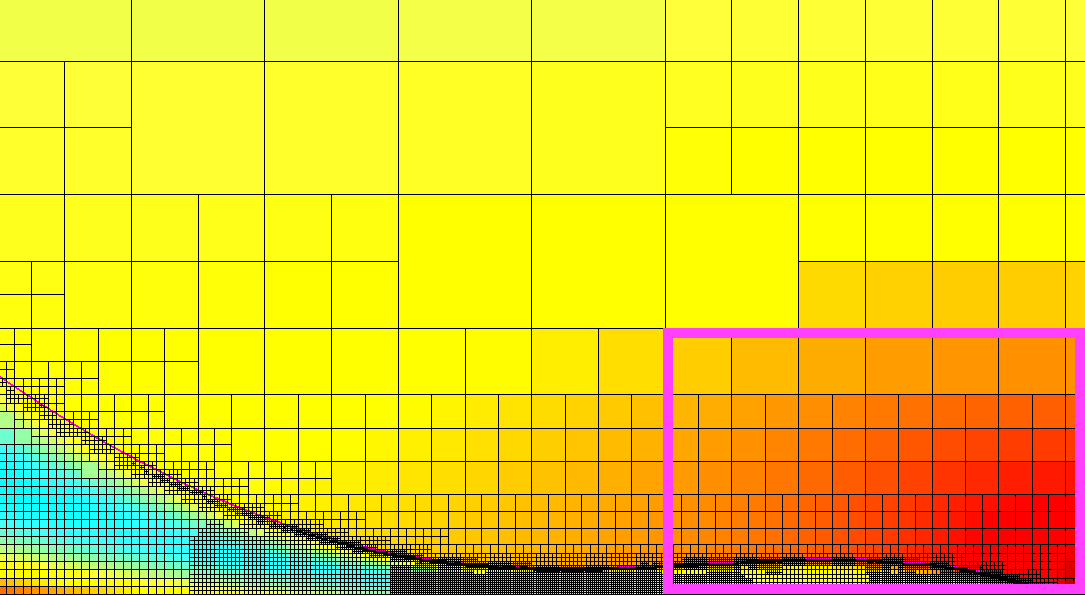}%
  
  \includegraphics[width=0.75\textwidth]{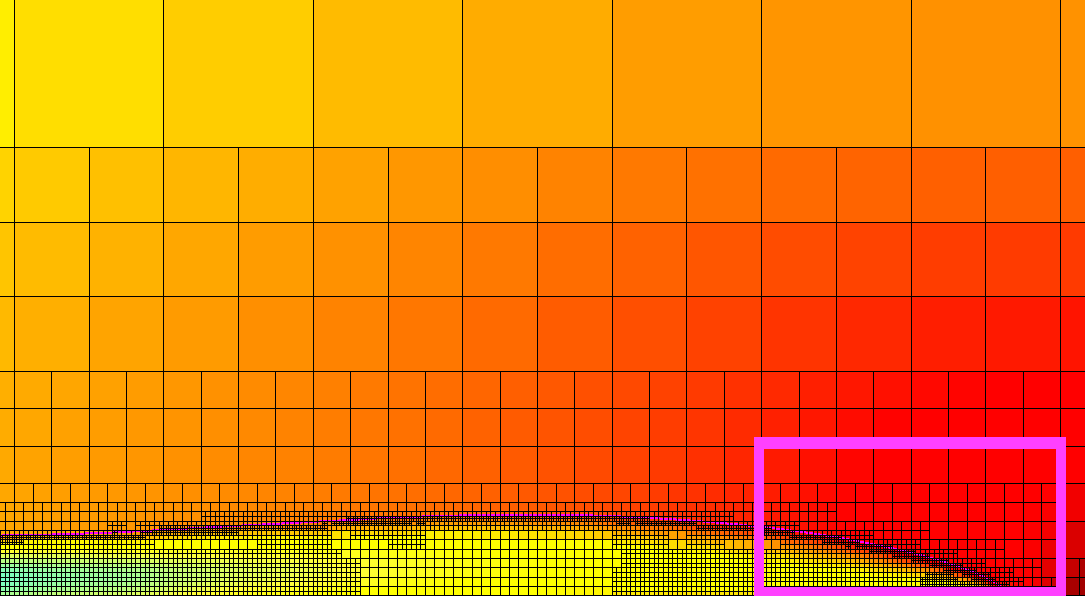}%
  
  \includegraphics[width=0.75\textwidth]{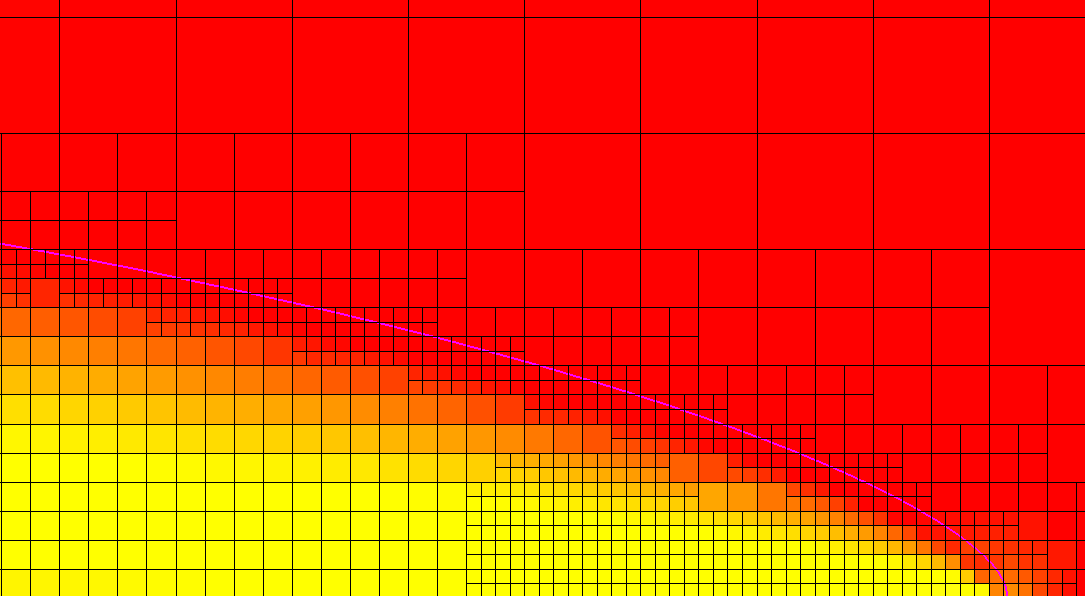}
        \label{nice2}
  \end{subfigure}
  \caption{Steady state solution of the reduced curtain coating system for Re = 30, Ca = 2.6, $\lambda = 10 \: \mu$m for 64 grid spacing per slip length. The color map corresponds to the x component of the velocity field. Left: foot of the curtain. Right: from top to bottom, successive zooms on the contact line.}
  \label{nice1}
\end{figure}

To validate our model, we reproduce the curtain coating configuration described in Liu et al. \cite{liu_vandre_carvalho_kumar_2016}. We consider a small curtain height $h_{c}$ = $10^{-2}$m and small curtain width $d_{c}$ = $10^{-3}$m with a large slip length $\lambda$ = $10^{-5}$m. As the accuracy of the interface reconstruction depends on the resolution of the smallest length scale, these previous considerations drastically decrease the computational cost of this multi-scale problem. The fluid properties are the following: $\rho_{l}$ = 1000 $kg.m^{-3}$, $\rho_{g}$ = 1.2 $kg.m^{-3}$, $\mu_{l}$ = 25 mPa.s, $\mu_{g}$ = 0.018 mPa.s and $\sigma$ = 70 mN.$m^{-1}$. The viscosity ratio $\mu_{g} / \mu_{l} = 7.2 \: 10^{-4}$ is kept constant in our computations. The substrate velocity $U$ is varied from 0.1 to 10 $m.s^{-1}$ and the feed flow velocity $V$ from 0.1 to 1 $m.s^{-1}$. The dimensionless numbers governing the flow are: the capillary number Ca $= \mu_{l} U /\sigma$ varying from 0.1 to 2.5, the Reynolds number Re $= \rho_{l} V d_c/ \mu_{l}$ varying from 1 to 40 and the Bond number Bo $= (\rho_{l} g / \sigma)(d_{c} V /U)^{2}$ varying from $10^{-3}$ to $10^{-1}$. The microscopic contact angle is kept constant, $\theta_{m} = 90^\degree$. In \textbf{Figure \ref{nice1}}, we show an example of a VOF simulation for a grid spacing $\Delta = 0.156 \: \mu$m, corresponding to 64 grid spacings per slip length. The adaptive mesh refinement allows a good resolution of the interface at the triple point. In this particular case, for Re = 30 and Ca = 2.6, a steady state solution is reached.

\begin{figure}[h]
  \centering
  \begin{subfigure}[h]{0.3\textwidth}
    \includegraphics[width=0.9\textwidth]{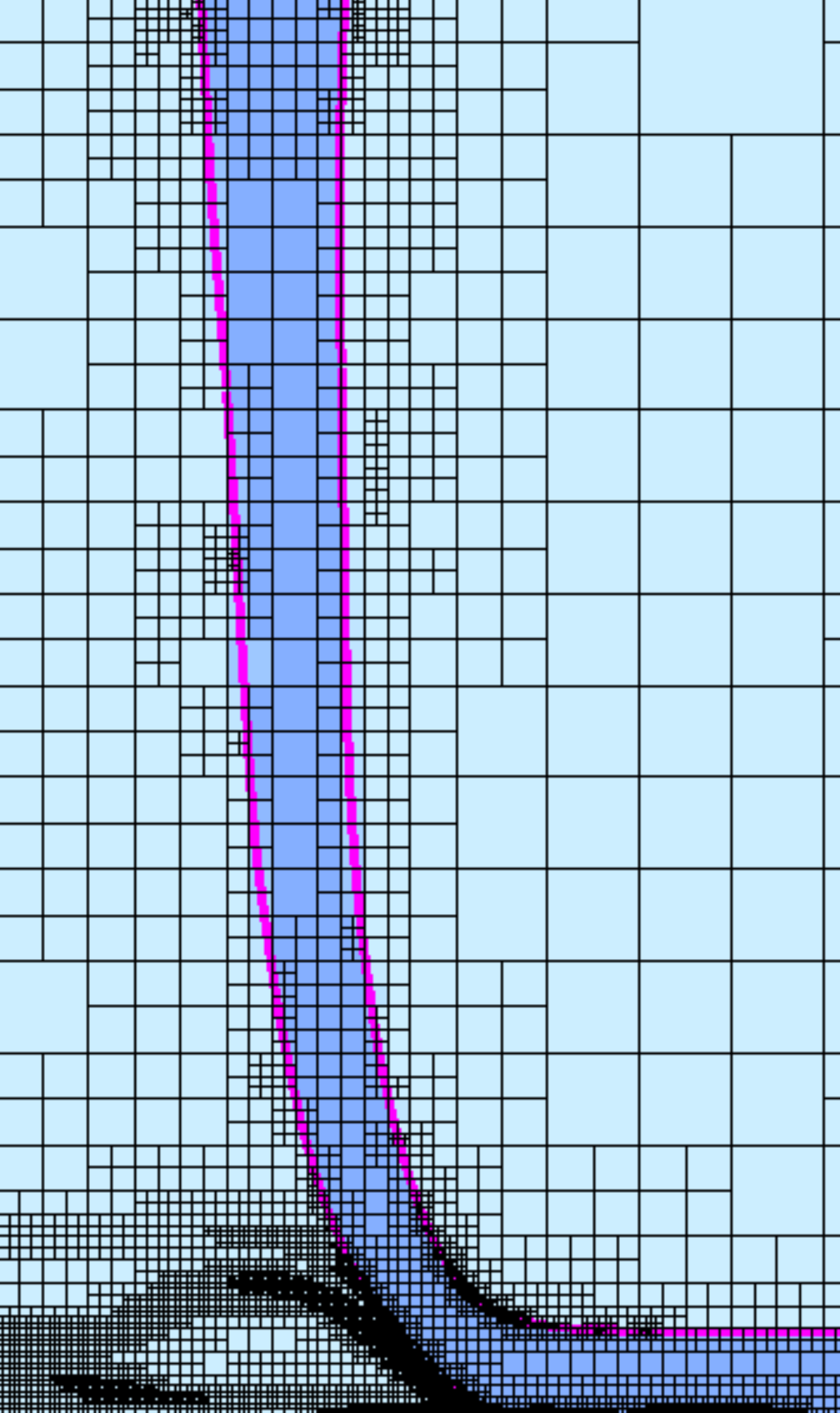}
  \end{subfigure}
  \begin{subfigure}[h]{0.3\textwidth}
    \includegraphics[width=0.9\textwidth]{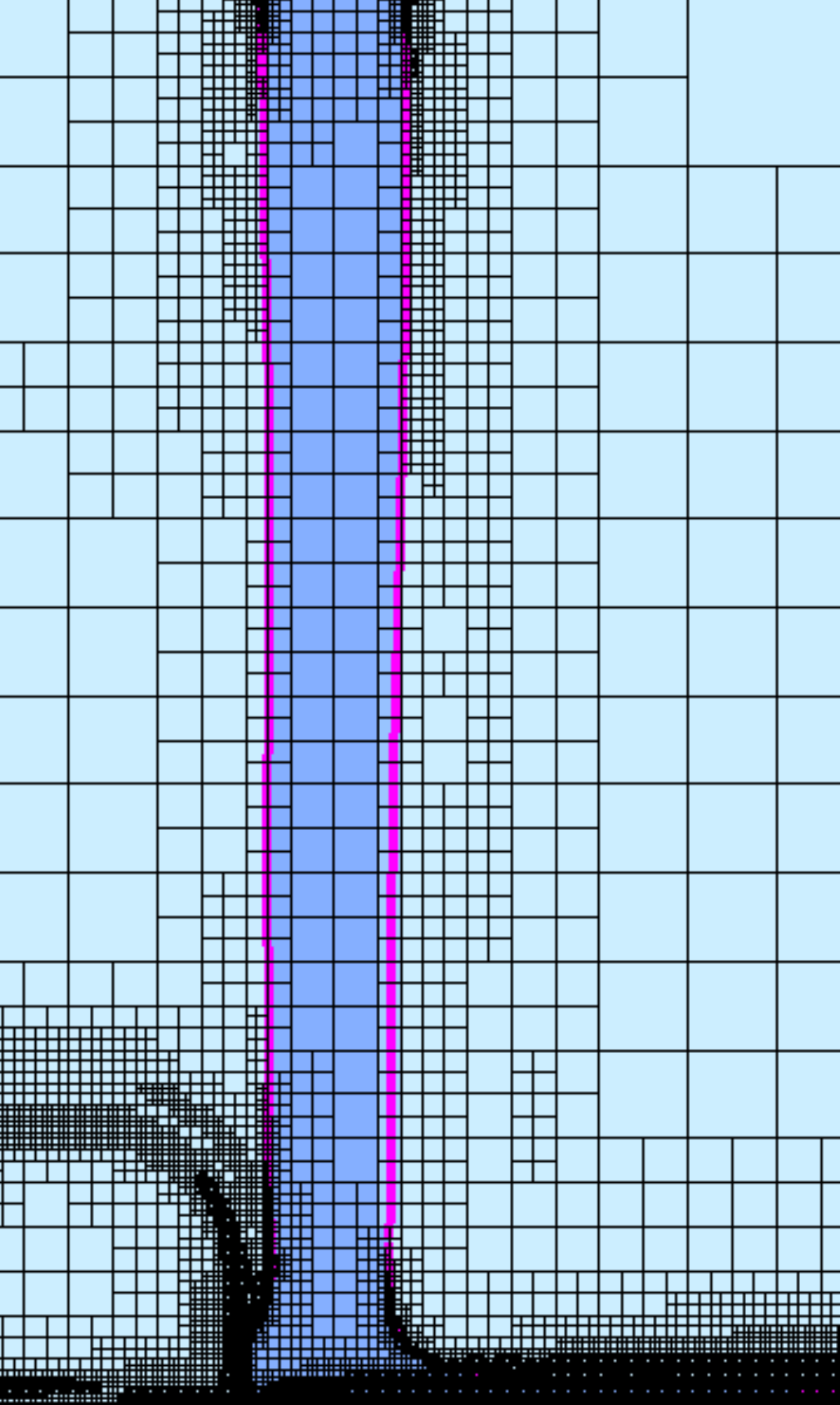}
  \end{subfigure}
  \begin{subfigure}[h]{0.3\textwidth}
    \includegraphics[width=0.9\textwidth]{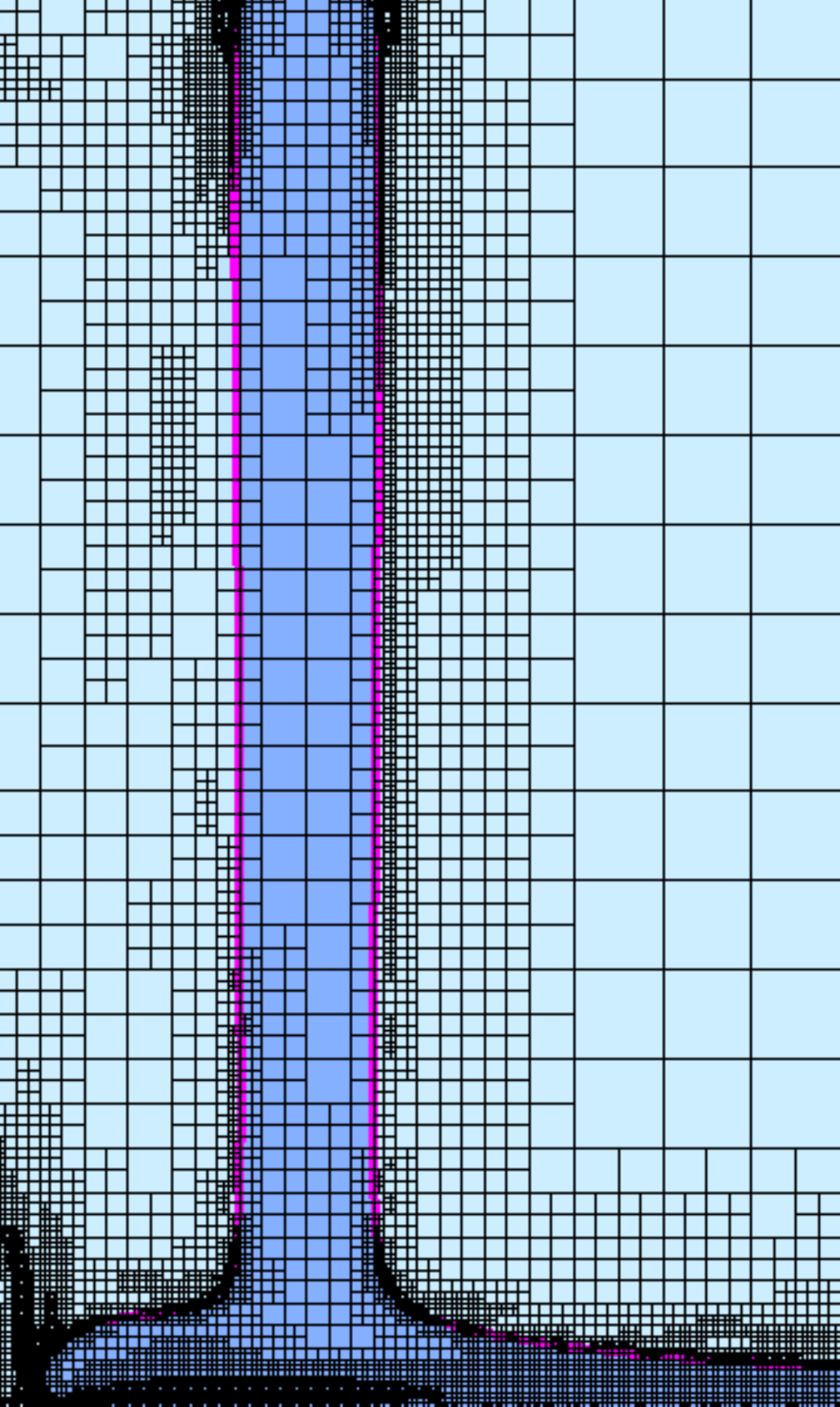}
  \end{subfigure}
  \caption[Example of interface resolution with adaptive mesh refinement for the curtain coating system.]{Steady state solution for the three different flow configurations. From left to right: bead pulling (Re = 12, Ca = 1.33), beneath the liquid curtain (Re = 30, Ca = 2.6), heel formation (Re = 35, Ca = 1.36).}
  \label{heel}
\end{figure}

\begin{figure}[h]
\begin{center}
    \includegraphics[width=0.85\textwidth]{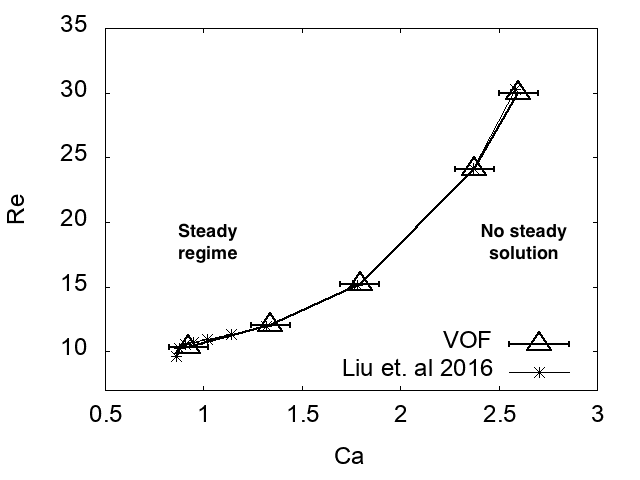}
    \caption{Stability analysis of the reduced curtain coating system and comparison with computations of Liu et al. \cite{liu_vandre_carvalho_kumar_2016}.}
    \label{liustab}
    \end{center}{}
\end{figure}

\begin{figure}[h]
  \centering
  \begin{subfigure}[h]{0.45\textwidth}
    \includegraphics[width=\textwidth]{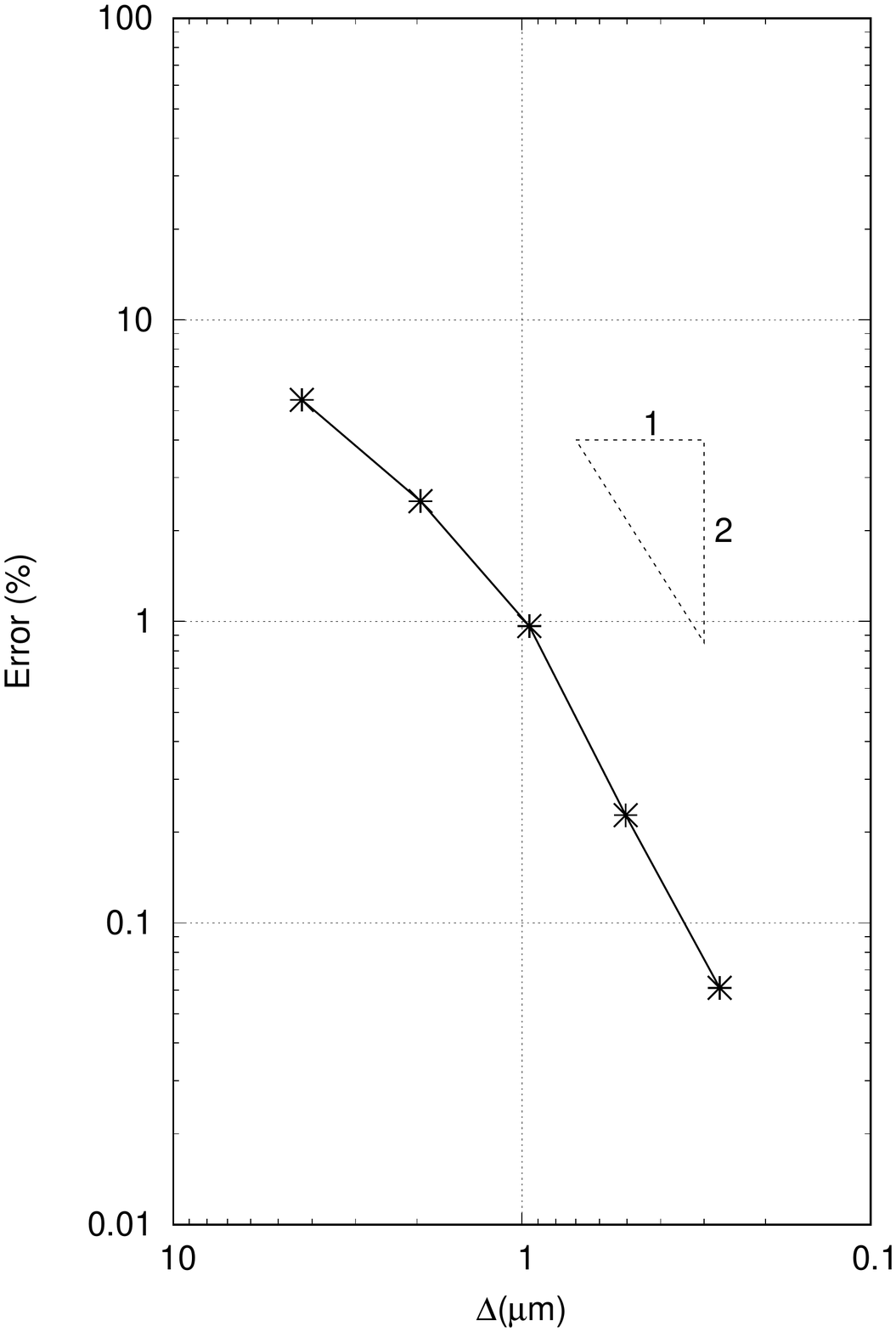}
    \caption{Error (\%) CL position.}
  \end{subfigure}
  \begin{subfigure}[h]{0.45\textwidth}
    \includegraphics[width=\textwidth]{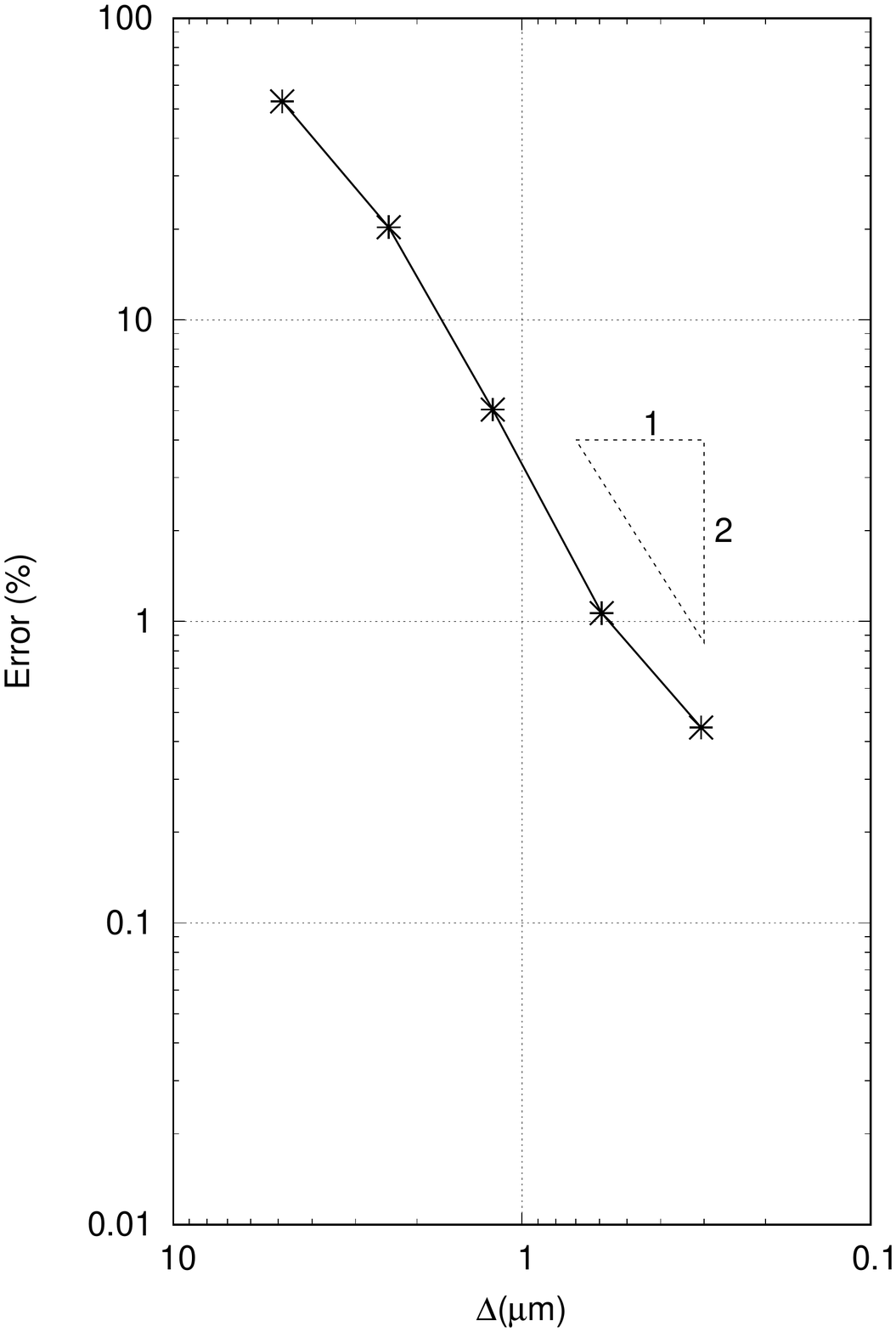}
    \caption{Error (\%) IP distance.}
  \end{subfigure}
  \caption{Convergence study of the reduced curtain coating system for Re = 30, Ca = 2.6 and $\lambda = 10  \: \mu$m. The reference solution is taken for 64 grid points per slip length.}
  \label{conv}
\end{figure}

To determine whether the simulation with a given set of physical parameters reaches the steady-state solution, we set a very large final time and compute, at each time step, the difference on the velocity field between two subsequent time steps. If the difference is lower than a given threshold, we can conclude that the flow has reached a steady state and that there is no wetting failure.
By varying the substrate velocity and the feed flow velocity, we are able to recover the same qualitative flow configurations: bead pulling, right beneath the liquid curtain and heel formation (\textbf{Figure \ref{heel}}) and a similar coating window (\textbf{Figure \ref{liustab}}) as in \cite{liu_vandre_carvalho_kumar_2016}. The hydrodynamic assist has the most impact (ie. the moving plate velocity $U$ is maximum) when the contact line is beneath the liquid curtain. This configuration allows a stronger pressure due to the liquid inertia at the triple point preventing the formation of bubbles and therefore preventing the wetting failure from occurring. To determine the coating window, we look for the first unsteady solution while increasing the Ca number for a given Re number. The error bar relates to the difference in Ca values between the last steady solution and the first unsteady one. A similar study will be performed when comparing to experiments in \textbf{Section 4}. A convergence study of the resolution of the interface is conducted for this configuration. As the maximal level of refinement is increased, the resolution of the interface at the contact line is improved. The microscopic contact angle tends to the prescribed one of 90° as the smallest cell size is decreased from 10 $\mu$m to 0.156 $\mu$m, corresponding to a number of grid points per slip length increasing from 1 to 64. We chose the 64 grid points per slip length solution as the reference solution for the contact line position. In \textbf{Figure \ref{conv}}, we plotted the relative error of the contact line position and the relative error of the distance from the contact line position to the inflexion point as a function of $\Delta$
for the Re = 30, Ca = 2.6 case, which is close to the stability limit. The results obtained for both quantities considered show a second-order convergence of the VOF method.
Moreover, the distance from the contact line position to the inflexion point of the reference solution is 50 $\mu$m. This result compares favorably with the experimental techniques used to compute the contact angle \cite{Blake1999,Blake2002}.

We have demonstrated that for a sufficient resolution of the interface, our VOF model with a Navier boundary condition on the moving substrate coupled to an imposed contact angle is able to reproduce the main stability results of the curtain coating configuration of Liu at al. \cite{liu_vandre_carvalho_kumar_2016}.

{}

\FloatBarrier

\section{Comparison with experiments}
\label{sec3}
We now propose to compare the results of the two-phase Navier-Stokes VOF model to the experimental observations of Blake et al. \cite{Blake1999} and Marston et al. \cite{Marston2009}.
For each series of experiments, by varying the substrate velocity $U$ and the feed flow velocity $V$, we look for the stability limit of the system. The boundary of the coating window corresponds to the critical substrate velocity at different liquid flow rates.

\subsection{Blake et al. (1999)}
\begin{figure}[h]
\begin{center}
    \includegraphics[width=0.85\textwidth]{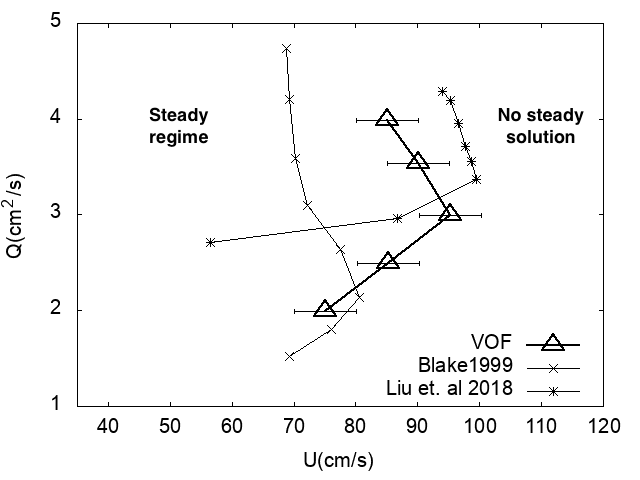}
    \caption[Stability analysis and comparison with experiments of Blake et al. and computations of Liu et al.]{Stability analysis and comparison with experiments of Blake et al. \cite{Blake1999} and computations of Liu et al. \cite{Liu2018}. The model parameters are: $\mu_{l}$ = 25 mPa.s, $\mu_{g}$ = 0.018 mPa.s, $\sigma$ = 64 mN.$m^{-1}$, $h_{c}$ = 3 cm, $\theta_{m} = 67^\degree$ and  $\lambda$ = $5$ nm.}
    \label{test}
    \end{center}{}
\end{figure}

In this section, we compare our model prediction with the experimental observations of Blake et al. \cite{Blake1999}. The values of experimental parameters of the system are taken as inputs for our simulations (curtain height, liquid viscosity, equilibrium surface tension, and imposed contact angle). The physical slip length used in \cite{Liu2018} is $\lambda = 5nm$. In this case, the slip length is not well-resolved as $\Delta = 230nm$ and there is an implicit numerical slip of $\Delta/2$. Therefore, the modification of the slip length doesn't affect the substrate velocity and induces no modification in the stability window in the numerics as the smallest $\Delta$ attainable is larger than $\lambda$. The stability limit curve of the VOF model is computed by interpolating between a stable and unstable solution, showed by the error bars, for a fixed feed flow rate $Q = V d_{c}$ (\textbf{Figure \ref{test}}). The numerical results show a maximum flow feed velocity $V$ of 3 $cm/s$ for a plate velocity $U$ of 90 cm/s whereas the experimental results from  \cite{Blake1999} give a maximum flow feed velocity of 2.15 cm/s for a plate velocity of 80.8 cm/s. The discrepancy between experimental observations and numerical results may be a consequence of the poor resolution of the smallest length scale in this case but also of the simplifications made at the boundary. The microscopic contact angle could depend on the parameters controlling the flow \cite{Blake2002,Shikhmurzaev1994,Wilson2006,Eggers2005}. A further regularization of the contact line singularity with a generalized Navier boundary condition \cite{quian2003}, where the contact angle is a dynamic one, might be needed. Even though the numerical results do not quantitatively match the experiments, we are still able to recover the shape of the stability-limit curve. The non-monotonic behaviour of the system is preserved.

{}

\subsection{Marston et al. (2006)}
\begin{figure}[h]
\begin{center}
    \includegraphics[width=0.85\textwidth]{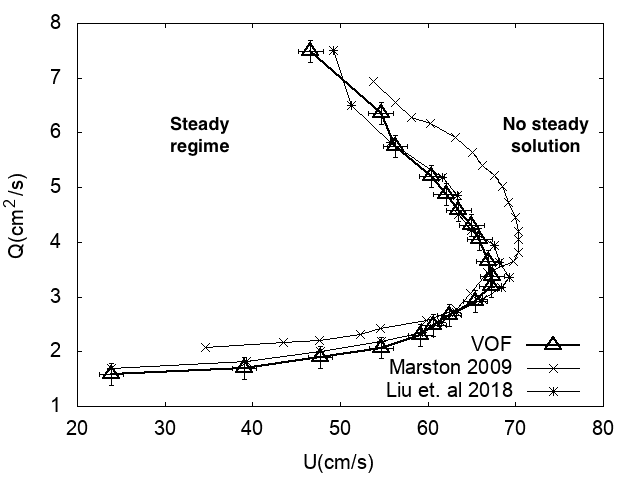}
    \caption[[Stability analysis and comparison with experiments of Blake et al. and computations of Liu et al.]{Stability analysis and comparison with experiments of Marston et al. \cite{Marston2009} and computations of Liu et al. \cite{Liu2018}. The model parameters are: $\mu_{l}$ = 117 mPa.s, $\mu_{g}$ = 0.018 mPa.s, $\sigma$ = 67 mN.$m^{-1}$, $h_{c}$ = 2.6 cm, $\theta_{m} = 67^\degree$ and $\lambda$ = $450$ nm.}
    \label{test2}
    \end{center}{}
\end{figure}

We now compare our model prediction with the experimental observations of Marston et al. \cite{Marston2009}. As with the previous comparison, we take the flow parameters as inputs for our system. In this configuration, $\Delta = 230nm $, so we are able to resolve the slip length determined in \cite{Marston2009} of the order of hundreds of nanometers. By increasing $\lambda$, the substrate velocity increases and the stability window is shifted towards the left, meaning that for the maximum substrate velocity the solution becomes unstable. We found that the best slip value was the one used in \cite{Liu2018}, $\lambda = 450nm$. As we can see in \textbf{Figure \ref{test2}}, the VOF model results are in good agreement with the experimental observations and almost perfectly match previous numerical computations. As noted in \cite{Liu2018}, the very large value of slip length found may come from the condition on the substrate. The substrate is pre-wetted in Marston et al. experiments \cite{Marston2009} whereas it is dry in our simulations.

{}
\FloatBarrier
\section{Conclusion}
In this work, we have solved the two-phase Navier-Stokes equations with a Navier boundary condition and a constant contact angle by a Volume-of-Fluid method on adaptive Cartesian meshes to predict the onset of wetting failure in a curtain coating system. We first computed a reduced curtain coating system, using the same configuration as Liu et al. \cite{liu_vandre_carvalho_kumar_2016}, in order to validate our model. We showed good agreement on the coating window and qualitative flow configurations with prior computations as well as a second-order convergence of the contact line position and the distance to the inflexion point. We are able to accurately represent the physics of the system and reproduce the non-monotonic behaviour of the substrate velocity with respect to the flow feed. Moreover, we compared our model with real experimental setups from Blake et al. \cite{Blake1999} and Marston et al. \cite{Marston2009} and new computations of Liu et al. \cite{Liu2018}. In the comparison with Blake et al. \cite{Blake1999}, although we were able to recover the shape of the stability limit, we were not able to match quantitatively the maximum feed flow velocity and maximum substrate velocity. This difference, which is similar to that in Liu et al. \cite{Liu2018} computations, may be a result of the poor resolution of the smallest length scale and the simplifications on the boundary condition at the solid boundary. In fact, the NBC coupled with a constant contact angle might not be enough to regularize the solution at the triple point. A further regularization, with a generalized Navier boundary condition (GNBC) \cite{quian2003}, where the contact angle is a dynamic one could help match the experimental observations.
Indeed, as it has been shown in Fricke et al. \cite{FRICKE201926} regular solutions for standard Navier Slip with a dynamic contact angle behave unphysically. The uncompensated Young's stress that is the extra term in the GNBC could relate to the evaporation phenomenon at the contact line (ie. phase change) that is absent in the sharp limit interface simulations. On the other hand, when comparing to Marston et al. \cite{Marston2009}, we found that the results matched quite well the experimental observations, provided one used the slip length as an adjustable parameter.

\appendix
\setcounter{figure}{0}    
\renewcommand{\theequation}{\thesection\arabic{equation}}
\renewcommand{\thefigure}{\thesection.\arabic{figure}}
\section{Details on the Navier-Stokes solver}
\label{app}
In this section we present some of the characteristics of the Volume-of-Fluid Navier-Stokes solver. For further details, we refer the reader to  \cite{Scardovelli1999,Popinet1999,Popinet2009,Popinet2015,Popinet2018}.  
\subsection{Temporal discretisation}
A staggered in time discretisation of the volume-fraction/density and pressure combined with a time-splitting projection method leads to the following time discretisation:

\begin{center}
    
$$\begin{array}{c} \rho_{n+\frac{1}{2}}\left[\dfrac{\mathbf{u}_{*}-\mathbf{u}_{n}}{\Delta t}+\mathbf{u}_{n+\frac{1}{2}} \cdot \mathbf{\nabla} \mathbf{u}_{n+\frac{1}{2}} \right] = \mathbf{\nabla}\cdot\left[\mu_{n+\frac{1}{2}}\left(D_{n}+D_{*}\right)\right]+\left(\sigma \kappa \delta_{s} \mathbf{n} \right)_{n+\frac{1}{2}} \\  
\\ \dfrac{c_{n+\frac{1}{2}}-c_{n-\frac{1}{2}}}{\Delta t}+\mathbf{\nabla} \cdot\left(c_{n} \mathbf{u}_{n}\right)=0
\\ 
\\ \mathbf{u}_{n+1}=\mathbf{u}_{\star}-\dfrac{\Delta t}{\rho_{n+\frac{1}{2}}} \mathbf{\nabla} p_{n+\frac{1}{2}}\end{array}$$

\end{center}{}
which requires the solution of the Poisson equation:
\begin{center}
    $$\begin{array}{c}\mathbf{\nabla} \cdot\left[\dfrac{\Delta t}{\rho_{n+\frac{1}{2}}} \mathbf{\nabla} p_{n+\frac{1}{2}}\right]=\mathbf{\nabla} \cdot \mathbf{u}_{\star}\end{array}$$
\end{center}{}
The momentum equation can be rewritten as: $$\begin{array}{l}{\dfrac{\rho_{n+\frac{1}{2}}}{\Delta t} \mathbf{u}_{\star}-\mathbf{\nabla} \cdot\left[\mu_{n+\frac{1}{2}} \mathbf{D}_{\star}\right]=\mathbf{\nabla} \cdot\left[\mu_{n+\frac{1}{2}} \mathbf{D}_{n}\right]+\left(\sigma \kappa \delta_{s} \mathbf{n}\right)_{n+\frac{1}{2}}+\rho_{n+\frac{1}{2}}\left[\dfrac{\mathbf{u}_{n}}{\Delta t}-\right.} \\ {\left.\mathbf{u}_{n+\frac{1}{2}} \cdot \mathbf{\nabla} \mathbf{u}_{n+\frac{1}{2}}\right]}\end{array}$$
where the right-hand side depends only on values at time $n$ and $n + 1/2$. This equation is solved using a multilevel Poisson solver. The velocity advection term $\mathbf{u}_{n+\frac{1}{2}} \cdot \mathbf{\nabla} \mathbf{u}_{n+\frac{1}{2}}$ is estimated using the Bell-Collela-Glaz second-order upwind scheme.\\

\subsection{Spatial discretisation}
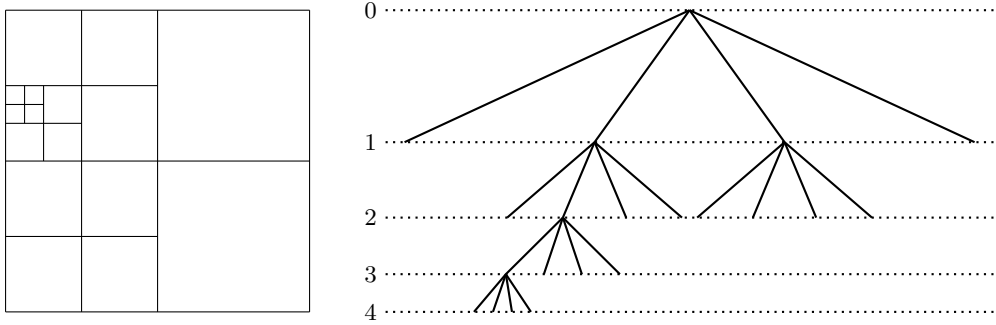
\begin{figure}[h]
\begin{center}
\begin{tikzpicture}
\draw [] (0,0) -- (4,0);
\draw [] (0,1) -- (2,1);
\draw [] (0,2) -- (4,2);
\draw [] (0,3) -- (2,3);
\draw [] (0,4) -- (4,4);

\draw [] (0,2.5) -- (1,2.5);
\draw [] (0,2.75) -- (0.5,2.75);

\draw [] (0,0) -- (0,4);
\draw [] (1,0) -- (1,4);
\draw [] (2,0) -- (2,4);
\draw [] (4,0) -- (4,4);

\draw [] (0.5,2) -- (0.5,3);
\draw [] (0.25,2.5) -- (0.25,3);

\draw [thick, dotted] (5,0) -- (13,0);
\draw [thick, dotted] (5,0.5) -- (13,0.5);
\draw [ thick, dotted] (5,1.25) -- (13,1.25);
\draw [ thick, dotted] (5,2.25) -- (13,2.25);
\draw [ thick, dotted] (5,4) -- (13,4);

\draw (5,0) node[left] {4};
\draw (5,0.5) node[left] {3};
\draw (5,1.25) node[left] {2};
\draw (5,2.25) node[left] {1};
\draw (5,4) node[left] {0};

\draw [thick] (9,4) -- (5.25,2.25);
\draw [thick] (9,4) -- (7.75,2.25);
\draw [thick] (9,4) -- (10.25,2.25);
\draw [thick] (9,4) -- (12.75,2.25);

\draw [thick] (7.75,2.25) -- (6.6,1.25);
\draw [thick] (7.75,2.25) -- (8.9,1.25);
\draw [thick] (7.75,2.25) -- (7.333,1.25);
\draw [thick] (7.75,2.25) -- (8.166,1.25);

\draw [thick] (10.25,2.25) -- (9.1,1.25);
\draw [thick] (10.25,2.25) -- (11.4,1.25);
\draw [thick] (10.25,2.25) -- (9.833,1.25);
\draw [thick] (10.25,2.25) -- (10.666,1.25);

\draw [thick] (7.333,1.25) -- (6.583,0.5);
\draw [thick] (7.333,1.25) -- (8.083,0.5);
\draw [thick] (7.333,1.25) -- (7.083,0.5);
\draw [thick] (7.333,1.25) -- (7.583,0.5);

\draw [thick] (6.583,0.5) -- (6.163,0);
\draw [thick] (6.583,0.5) -- (6.913,0);
\draw [thick] (6.583,0.5) -- (6.413,0);
\draw [thick] (6.583,0.5) -- (6.663,0);
\end{tikzpicture}
\end{center}

    \caption{Example of quadtree discretisation and corresponding tree representation.}
    \label{quad}
\end{figure}

Space is discretised using a quadtree partitioning in 2D (\textbf{Figure \ref{quad}}).

All the variables are collocated at the centre of each square discretisation volume. Consistently with a finite-volume formulation, the variables are interpreted as the volume-averaged values for the corresponding discretisation volume. A projection method is used for the spatial discretisation of the pressure correction equation and the associated divergence in the Poisson equation.

\subsection{Volume-of-Fluid advection scheme}
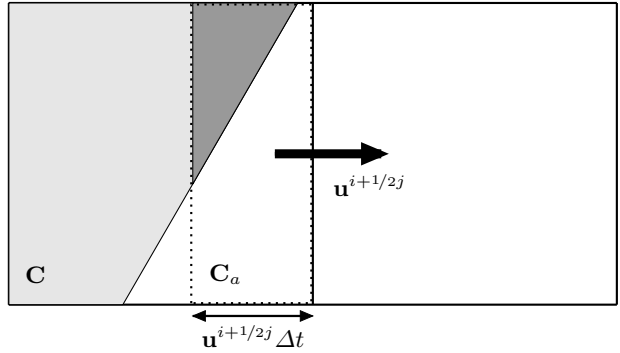
\begin{figure}[h]
\begin{center}
\begin{tikzpicture}

\draw [thick] (0,0) -- (8,0);
\draw [thick] (0,0) -- (0,4);
\draw [thick] (8,0) -- (8,4);
\draw [thick] (0,4) -- (8,4);

\draw [thick] (4,0) -- (4,4);
\draw [fill=gray!20] (1.5,0) -- (3.8,4) -- (0,4) -- (0,0) -- (1.5,0);
\draw [fill=gray!80] (2.42,1.6) -- (3.8,4) -- (2.42,4) -- (2.42,1.6);
\draw [thick, dotted] (2.4,0.02) rectangle (3.98,3.98);

\draw[
        triangle 45 -triangle 45,
        line width=0.01mm,
        postaction={draw, line width=0.03cm, shorten >=0.1cm, -}
    ] (2.42,-0.15) -- (4,-0.15);
\draw[
        -triangle 45,
        line width=0.5mm,
        postaction={draw, line width=0.12cm, shorten >=0.1cm, -}
    ] (3.5,2) -- (5,2);
\draw (3.21,-0.15) node[below] {$\mathbf{u}^{i+\nicefrac{1}{2}j} \Delta t$};

\draw (4.75,1.85) node[below] {$\mathbf{u}^{i+\nicefrac{1}{2}j}$};

\draw (0.1,0.4) node[right] {$\mathbf{C}$};
\draw (2.52,0.4) node[right] {$\mathbf{C}_a$};
\end{tikzpicture}
\end{center}
  \caption{Example of geometrical flux estimation.}
      \label{vof}

\end{figure}

To solve the advection equation the geometrical VOF scheme is used and proceeds in two steps:
\begin{enumerate}
    \item Interface reconstruction.
    \item Geometrical of flux estimation and interface advection.
\end{enumerate}{}
The reconstruction is a ‘piecewise linear interface calculation’ (PLIC), followed by a Lagrangian advection. In the PLIC technique, given a volume fraction $c(\mathbf{x},t)$ and an approximate normal vector $\mathbf{n}$, a linear interface is constructed within each interface cell, which corresponds exactly to $c(\mathbf{x},t)$ and $\mathbf{n}$. In  \textbf{Figure \ref{vof}}, we illustrate the principle of geometrical flux estimation. The total volume which will be fluxed to the right-hand neighbour is delimited with a dashed line. The fraction of this volume occupied by the first phase is indicated by the dark grey triangle.

\bibliographystyle{unsrt}
\bibliography{biblist}


\end{document}